\documentclass{article}
\usepackage{ijcai23}

\usepackage{times}
\usepackage{soul}
\usepackage{url}
\usepackage[hidelinks]{hyperref}
\usepackage[utf8]{inputenc}
\usepackage[small]{caption}
\usepackage{graphicx}
\usepackage{amsmath}
\usepackage{amsthm}
\usepackage{booktabs}
\usepackage{algorithm}
\usepackage{algorithmic}
\usepackage[switch]{lineno}
\usepackage{paralist}
\usepackage{multirow}
\usepackage{caption}
\usepackage{subcaption}

\title{McUDI: Model-Centric Unsupervised Degradation Indicator for Failure Prediction AIOps Solutions}

\author{
Lorena Poenaru-Olaru$^1$
\and
Luis Cruz$^1$\and
Jan Rellermeyer$^{2}$\And
Arie van Deursen$^1$
\affiliations
$^1$Software Engineering, TU Delft, Netherlands\\
$^2$Dependable and Scalable Software Systems, Leibniz University Hannover, Germany\\
\emails
\{L.Poenaru-Olaru, L.Cruz, arie.vandeursen\}@tudelft.nl,
rellermeyer@vss.uni-hannover.de
}

\begin{document}

\maketitle

\begin{abstract}
Due to the continuous change in operational data, AIOps solutions suffer from performance degradation over time. Although periodic retraining is the state-of-the-art technique to preserve the failure prediction AIOps models' performance over time, this technique requires a considerable amount of labeled data to retrain. In AIOps obtaining label data is expensive since it requires the availability of domain experts to intensively annotate it. In this paper, we present McUDI, a model-centric unsupervised degradation indicator that is capable of detecting the exact moment the AIOps model requires retraining as a result of changes in data. We further show how employing McUDI in the maintenance pipeline of AIOps solutions can reduce the number of samples that require annotations with 30k for job failure prediction and 260k for disk failure prediction while achieving similar performance with periodic retraining.
\end{abstract}

\section{Introduction}

%Paragraph about AIOps \& importance
Given the large amount of operational data generated by large-scale software systems, manual inspection for abnormalities or failures in the systems becomes impractical. This led to the appearance of the AIOps (Artificial Intelligence for IT Operations) research field. AIOps is a specific machine learning (ML) application that applies machine learning techniques to operational data to identify or predict failures or anomalies that occur in a large-scale software system. 

As observed in other real-world ML applications, such as healthcare, manufacturing, finance, and agriculture~\cite{aiaging}, one of the biggest challenges in building AIOps models is the temporal quality degradation (AI aging)~\cite{aiopschallenges},~\cite{datasplittingdecisions},~\cite{towardsaconsistentinterpretation}. The model quality (performance/accuracy) degradation is a consequence of the evolving character of operational data, also known as concept drift. In some situations, concept drift leads to severe model degradation~\cite{BAYRAM2022108632}, which impacts the reliability and trustworthiness of AIOps models. Moreover, a considerable degradation in the quality of AIOps models generates service interruptions~\cite{aiopschallenges}, which causes substantial monetary losses (between \$300,000 and \$400,000 according to~\cite{towardsaconsistentinterpretation}).

To overcome the challenge of model performance degradation, previous work proposes to periodically retrain (update) AIOps models~\cite{datasplittingdecisions},~\cite{towardsaconsistentinterpretation}. However, in a practical scenario, periodic retraining is sometimes unfeasible due to the hidden deployment costs~\cite{haakman2021ai}. Once a model is updated, it undergoes compliance verification, which is time-consuming~\cite{haakman2021ai}, and afterward, engineers need to ensure its smooth integration with the software system that incorporates the AIOps model~\cite{monitoringDataDistribution}. Thus, a solution that reduces the number of necessary updates without harming the performance of the model is desired.

Previous work~\cite{modelmaturity} proposed a more systematic solution to periodic model retraining, namely retraining only when necessary. This solution involves updating an AIOps model when a model degradation indicator detects a significant drop in its performance. However, this solution requires having true labels available in real-time, which implies that operational engineers continuously manually annotate data, making it unfeasible for real-world scenarios. Thus, a label-independent model degradation indicator is required.

%is a major limitation 

%because collecting high-quality data is expensive, as operational engineers have to continuously manually annotate data (e.g., system anomalies). % to continue

%Although the systematic solution presented in~\cite{modelmaturity} is promising since it lowers the retraining frequency, it comes with a high cost of obtaining high-quality labels. Thus, the same solution would be feasible if the model degradation indicator would not be dependent on the immediate availability of labels. %smooth transition to data-based techniques +model centric

To overcome these limitations, we propose McUDI a novel model-centric label-independent performance degradation indicator, which indicates when an AIOps failure prediction model requires retraining. McUDI computes the feature importance ranking from the trained AIOps failure prediction model and selects the most important features according to the average importance rank. To identify drift, McUDI applies the Kolmogorov-Smirnov (KS) statistical test on the data distribution computed from the most important features. Our contributions are the following:
\begin{compactenum}
    \item We propose a model-centric unsupervised failure prediction AIOps models degradation indicator and evaluate it on two operational datasets.
    \item We propose a maintenance pipeline that incorporates McUDI which reduces the need for high-quality labels necessary for retraining while preserving the performance of AIOps solutions.
    \item We evaluate the capability of the state-of-the-art data monitoring practices to identify performance degradation on failure prediction data.
    \item We publicly share our reproduction package\footnote{Replication Package: \url{https://github.com/LorenaPoenaru/aiops_failure_prediction}}.
\end{compactenum}

\section{Related Work}
\subsection{Concept Drift Definition and Evaluation}
In literature, the term concept drift generally refers to changes in the data that occur over time~\cite{BAYRAM2022108632}. Concept drift detectors are monitoring tools developed to capture these changes and raise alerts when they occur~\cite{learningUnderConceptDrift}. Plenty of drift detectors were proposed for classification problems~\cite{BAYRAM2022108632}. Some of the existing concept drift detection techniques monitor the changes in accuracy over time and they are, therefore, label-dependent, while others identify drifts only in the characteristics of the dataset, such as the data distribution~\cite{learningUnderConceptDrift}. Although the label-dependent drift detection techniques are reliable model performance degradation indicators, they are expensive to employ in real-world scenarios since true labels are not immediately available. Thus, a better alternative to real-world applications is the usage of data distribution-based drift detectors, which do not rely on labels. However, previous work~\cite{mypaper} has shown that these types of detectors are quite sensitive to small changes in the data that do not necessarily affect the accuracy of ML models.

When it comes to evaluation, concept drift detectors are evaluated in two manners: \textit{assessing the drift detection accuracy} and \textit{assessing the performance of a machine learning model over time}. The former assessment technique measures drift detectors' capability to correctly label drifts and non-drifts. Given that in real-world scenarios the moment when drift occurs is difficult to determine, this technique is employed when experimenting with synthetic data where the drift occurrence is fixed~\cite{learningUnderConceptDrift},~\cite{mypaper}. However, previous work presented a technique to label testing batches into drift and non-drift respectively using a Z test on the error-rate~\cite{datasplittingdecisions}. The latter assessment technique is usually preferred in real-world cases since knowledge regarding the moment of drift occurrence is not required ~\cite{BAYRAM2022108632}. In this assessment technique, to understand the impact of the detector, the performance (e.g. accuracy, ROC-AUC, etc) of an ML model that is retrained based on a drift detector is compared to the performance of a model that is never retrained~\cite{learningUnderConceptDrift}.

%It is primarily assessed using synthetic datasets since the moment of concept drift occurrence can be fixed while generating the data~\cite{learningUnderConceptDrift},~\cite{mypaper}. Usually detecting concept drift in real-world data is difficult to assess due to the lack of ground truth regarding the moment of drift occurrence. Another evaluation technique of drift detectors is monitoring the performance of a model over time when being retrained after a drift is detected vs not being retrained at all~\cite{BAYRAM2022108632}. This evaluation technique is usually preferred since the knowledge of the exact moment the concept drift occurs is not necessary and the drift detector is solely assessed according to its capability to preserve the performance of the ML model over time.

\subsection{AIOps Models Degradation due to Concept Drift}

AIOps models positively influence software delivery acceleration and software systems quality and security~\cite{aiopschallenges}. Despite their potential, it has been demonstrated that the performance of these models is not generalizable over time due to concept drift. AIOps models developed for defect prediction trained on current data cannot generalize well on future data~\cite{conceptdriftaffecteddefectdetection}. Operational data is constantly changing/evolving (concept drift occurs) due to uncontrollable factors such as user workloads or hardware/software upgrades. Concept drift on operational data negatively impacts the failure prediction AIOps models performances, resulting in an increase in their error rates~\cite{datasplittingdecisions},~\cite{towardsaconsistentinterpretation},~\cite{nodefailurepredictionconceptdrift}.

%Previous work demonstrated that concept drift occurs in operational data used to train AIOps failure prediction models (disk failure and job failure prediction) by identifying significant changes in the error rate of the model over time. 

Previous work highlights the importance of periodically updating AIOps models to mitigate the effect of concept drift and the necessity of monitoring AIOps models against concept drift~\cite{datasplittingdecisions},~\cite{towardsaconsistentinterpretation}. Research has been done on the effects of retraining failure prediction models based on a degradation indicator monitoring tool that requires labels~\cite{modelmaturity}. However, there is currently no research on unsupervised model degradation indicators. 

%However, periodic updates imply that high quality true labels are available immediately to retrain the AIOps model, which in some situations is hard to achieve since some AIOps applications imply that experts need to annotate the data to obtain labels \textbf{CITE SMTH}.

% talk about job and disk

\subsection{Data Monitoring}
Model monitoring refers to a set of tests performed after the model is deployed into production~\cite{monitoringMLM}. One essential part of monitoring machine learning models is data monitoring against concept drift.

One data monitoring technique is monitoring the number of features that become skewed over time~\cite{modelmonitoringgoogle}. This technique checks whether one feature in the training data behaves significantly differently than the same feature in the testing data and was previously used to assess data quality~\cite{monitoringFeatureChange}. Another popular data monitoring technique is continuously checking if the distribution of data changes over time~\cite{monitoringMLM},~\cite{modelmonitoringgoogle},~\cite{monitoringDataDistribution3},~\cite{monitoringDataDistribution4}. A change in the data distribution between training and testing implies that the model needs to be rebuilt~\cite{monitoringDataDistribution} or retrained with newer samples~\cite{monitoringDataDistribution2}. Although these monitoring techniques have been proposed, their capability of identifying the degradation of AIOps models has not yet been assessed. 

%\subsection{Change Detection}

\section{Proposed Method}
\subsection{Background Feature Importance}
Feature importance (FI) analysis has been used previously as a technique to explain the output of a certain machine learning model in both scientific and industrial settings~\cite{featureImportance1},~\cite{featureImportance2},~\cite{mdiFeatureImportanceRandomForests}. The FI is a feature ranking technique that shows how much each feature contributes to the predictive capability of a classifier~\cite{featureImportance3}. 

According to~\cite{mdiFeatureImportanceRandomForests} although the FI ranking computation is beneficial for understanding the behavior of classifiers, there is a high interest in developing techniques suitable for specific classifiers such as Random Forests~\cite{Breiman2001RandomF} and other tree-based classifiers~\cite{treebasedFI}. The reason for this is wide applicability within different research areas of tree-based classifiers. Thus the most common metric to compute the FI ranking from tree-based classifiers is the mean decrease in impurity (MDI)~\cite{mdi1},~\cite{mdi2},~\cite{mdi3},~\cite{mdiFeatureImportanceRandomForests},~\cite{treebasedFI}, also known as Gini importance. For each feature, this technique calculates the total decrease in impurity (loss) computed for each random split~\cite{mdiFeatureImportanceRandomForests}.
\subsection{McUDI}
McUDI is derived from one of the most commonly used unsupervised techniques to detect drift the Kolmogorov-Smirnov (KS) statistical test. This test is preferred over others since it is non-parametric, label-independent~\cite{conceptdriftadaptation}, has low computational costs~\cite{kslowcomputation}, and does not require a manually adjusted drift threshold~\cite{learningUnderConceptDrift}. The KS test is usually applied to assess whether the data distribution in the training set is different than the one in the testing set. Despite its popularity, one major limitation of KS is the high number of false alarms caused by the fact that the data distributions are computed using all the available features, while any change does not influence the model's performance in the data~\cite{mypaper}. McUDI aims to overcome this limitation by combining a technique of extracting the model's most important features with the KS drift detection method. The major difference between McUDI and KS is the fact that while KS computes the distribution using all available features, McUDI computes the distribution using solely the most important features, being, thereby, a model-centric unsupervised drift detector. McUDI is composed of 3 important steps, namely \textbf{extract important features}, \textbf{compute distributions}, \textbf{statistical test significance}.\\
\textbf{Extract Important Features.} 
We use the mean decrease in impurity to rank features according to their importance to the trained model. We extract the average importance rank among all features and consider the features with feature importance rank higher than the average relevant and discard the rest. Considering the average importance instead of a fixed percentage of important features increases the generalizability of McUDI since some models might consider 80\% of the features important, while others might consider only 50\%.\\
\textbf{Compute the Distributions.} After the less important features are filtered out, the data distribution needs to be estimated. Thus, in this step, the distribution of the training data and the distribution of the testing data on the most important features is computed.\\
\textbf{Statistical Test Significance.} The KS statistical test is employed to assess the similarity/dissimilarity between the distributions of the most important features from the training and testing data. The null hypothesis of this test is that the two distributions are similar. An alarm is raised (drift is detected) when the p-value is smaller than 0.05 (95\% confidence interval). 

\section{Research Questions}
This section highlights the research questions we aim to answer in our study.
\begin{compactenum}
\item How well do state-of-the-art (SOTA) concept drift monitoring techniques identify model degradation in failure prediction AIOps solutions (job \& disk failure)? 
\item How does our proposed model-centric approach (McUDI) compare to traditional SoTA monitoring practices?
\begin{compactenum}
\item How does McUDI compare to traditional SOTA monitoring practices in terms of drift detection accuracy?
\item How does McUDI compare to traditional SOTA monitoring practices in terms of model performance over time?
\end{compactenum}
%Would the model-centric approach improve the SoTA monitoring practices in detecting model degradation?
\item What is the effect of including McUDI in the maintenance pipeline of failure prediction AIOps solutions in terms of label costs compared to existing practices?
\end{compactenum}

%RQ1: How do sota monitoring practices %compare in AIOps use cases?
%RQ2: Would the model-centric approach %improve sota monitoring practices?
%RQ3: The impact of updating a model based on the indication of the monitoring tool.

%Rq3.1. impact on no retrainings
%Rq3.2 impact on no retrainings and cost of labeling
\section{Evaluation Methodology McUDI}
In this section, we present the employed operational datasets, the baseline methods, and performance evaluation. 
\subsection{Data \& Failure Prediction AIOps Model}
\subsubsection{Data}
For our study, we employ two open-source AIOps datasets for which previous work has shown that concept drift occurs~\cite{datasplittingdecisions}. 

The first dataset is the \textbf{Google Cluster Traces Dataset}~\cite{googleclusterdata} and contains information about traces extracted from real-world large cluster systems. This data was collected for 29 days (May 2011). In total, there are around 625K samples. Previous work~\cite{datasplittingdecisions},~\cite{googletraceprediction1} used this dataset to design job failure prediction models. 

The second dataset is the \textbf{Backblaze Disk Stats Dataset}~\cite{backblazedata} and contains information about HDD hard drive stats from 2013. Similar to previous work~\cite{datasplittingdecisions},~\cite{towardsaconsistentinterpretation} we are using 12 months of data corresponding to the year 2015, which contains around 7M samples of data. This dataset was used to design disk failure prediction models~\cite{diskfailureprevwork},~\cite{datasplittingdecisions},~\cite{towardsaconsistentinterpretation},~\cite{diskfailureprevwork1}. \\
\textbf{Data Preprocessing and Feature Selection} \\
To build the AIOps model that predicts disk failure, we employ 19 features out of which some of them have values accumulated over time, such as the ``reallocated sectors coun'' features, while others contain non-accumulated values over time, such as ``read error rate''. When building the AIOps model that predicts job failure, we employ a total of 15 features, out of which 6 are temporal features, such as the ``standard deviation of CPU'', and 9 are configuration features, such as ``memory and disk space''. 

% Scaling discussion
Unlike the original approach presented in~\cite{datasplittingdecisions}, we use a slightly different scaling method. Instead of fitting the scaler on the entire dataset and therefore transforming it, we use the scaler fitted on the first period of data only to transform the datasets from both the first and the second period. We argue that this data scaling method is more realistic for model deployment because using the whole dataset in this context would assume that we have access to future data.

%\subsection{Evaluation McUDI}
\subsubsection{Failure Prediction AIOps Model}
We build the failure prediction AIOps model using a Random Forests Classifier. We specifically choose a tree-based classifier since these types of classifiers are very well researched in terms of feature importance ranking extraction~\cite{featureImportance1},~\cite{featureImportance2},~\cite{featureImportance3} which is an important step in McUDI. Out of the tree-based classifiers we specifically choose Random Forests since it is a commonly used tree-based classifier in failure prediction AIOps solutions~\cite{datasplittingdecisions},~\cite{towardsaconsistentinterpretation},~\cite{diskfailureprevwork1},~\cite{googletraceprediction1}, ~\cite{diskfailureprevwork},~\cite{modelmaturity}. The hyperparameters of the Random Forests classifier are chosen according to previous work~\cite{datasplittingdecisions}.

\subsection{Baseline Model Degradation Indicators}
We begin our experiments by understanding how well the two data monitoring techniques, monitoring the number of features that change and monitoring changes in data distribution, capture model performance degradation. When monitoring the number of changing features we count the number of features that are different from training and testing using the KS statistical test. We further assess how well KS captures drift. %We further include in our evaluation baselines the most commonly used technique to mitigate the effects of concept drift on AIOps data, namely periodic retraining. 

%We further compare the accuracy of KS with the accuracy of McUDI in detecting performance degradation.

\subsection{Performance Evaluation}

To evaluate McUDI we employ the two aforementioned concept drift evaluation techniques, the drift detection accuracy assessment, and the performance over time of failure prediction model assessment.

%AIOps failure prediction model performance indicator. 

%Since the most commonly used technique to mitigate the effects of concept drift on AIOps data is periodically updating the model we do consider it as our upper baseline in terms of the best performance that can be achieved. Our lower baseline corresponds to the model performance generated by a model that was never updated. The goal is that by using a model degradation indicator the performance is considerably close to the upper baseline with a lower number of needed retraining times.

\subsubsection{Evaluation based on Drift Detection Accuracy}
To evaluate drift detectors based on their accuracy in correctly distinguishing between drifts and non-drifts, we first need to obtain the ground truth in terms of which testing batch is a drift and which one is a non-drift.
\par{\textbf{Extracting the Ground Truth.}} To extract the ground truth regarding when concept drift occurs we follow the same technique as previous work~\cite{datasplittingdecisions}. This technique implies that machine learning data is trained on data corresponding to the previous time period and evaluated on data from the current time period. The assumption is that the performance of the model on the training data (past data) is not statistically different than the performance of the model on the testing data (current data). If this assumption does not hold, then there is concept drift. In our paper, we assess the performance of the model on training and testing by computing the prediction error rate on training and testing, respectively, similar to previous work~\cite{datasplittingdecisions}.

\begin{figure}
    \centering
    \includegraphics[width=0.5\textwidth]{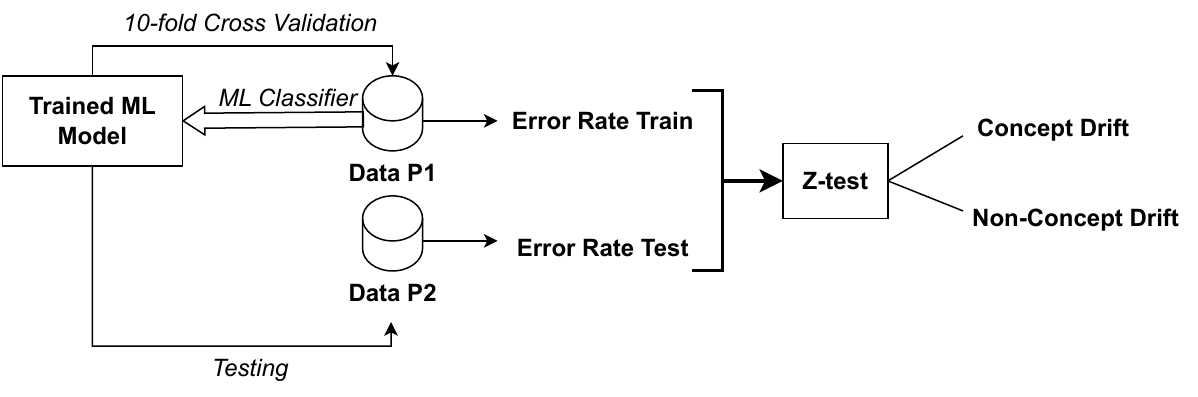}
    \caption{Obtaining the ground truth. Pipeline to assess the presence of concept drift between two batches (Data P1 and Data P2).}
    \label{figure:pieline_extracting_ground_truth}
    \vspace{-1em}
\end{figure}

In Figure \ref{figure:pieline_extracting_ground_truth} we depict the pipeline of identifying whether concept drift occurs between datasets extracted from two different periods, P1 and P2. We train an ML model using the data from the first period (P1) and test it on the data from the second period (P2). As it can be observed from Figure \ref{figure:pieline_extracting_ground_truth} we compute the error rate on training by performing a 10-fold cross-validation on Data P1 and the error rate on testing by testing the model on Data P2. On the two error rates, we apply a two-proportion Z-test to assess whether there is concept drift between the datasets from two different periods or not. The formula of this test is the following:
\begin{equation}\label{eqztest}
Z = \frac{\epsilon_{test} - \epsilon_{train}}{\sqrt{\epsilon(1-\epsilon)(\frac{1}{n_{train}}+\frac{1}{n_{test}})}}
\end{equation}

where $\epsilon_{train}$ is the prediction error rate on the training set, $\epsilon_{test}$ is the prediction error rate on the testing set, $\epsilon$ is the overall prediction error rate, $n_{train}$ is the length of the training set and $n_{test}$ is the length of the testing set.

The null hypothesis of the Z-test is that there is no significant difference between the machine learning model's performance on datasets extracted from the two different periods, implying that there is no concept drift. The null hypothesis is rejected when the p-value of the Z-test is lower than 0.05. The severity of concept drift is calculated as (($\epsilon_{test}$-$\epsilon_{train})/\epsilon_{train})$.

\par{\textbf{Assessing the Drift Detection Accuracy.}} Given that in each dataset, the ratio between the number of drifts and non-drifts is highly imbalanced, we employ three metrics that take into account the imbalance between the two, namely the \textit{Balanced Accuracy}, the \textit{Specificity} and the \textit{Sensitivity}. The Specificity shows how many drifts are correctly identified out of the total number of drifts. The Sensitivity shows how many non-drifts are correctly identified out of the total number of non-drifts. The Balanced Accuracy is the arithmetic mean between the Sensitivity and Specificity and it shows how many drifts and non-drifts are overall correctly classified. 

\begin{equation}\label{specificity}
Specificity = \frac{TN}{TN+FP}
\end{equation}; where \textit{TN} corresponds to True Negatives and FP corresponds to False Positive

\begin{equation}\label{sensitivity}
Sensitivity = \frac{TP}{TP+FN}
\end{equation}; where \textit{TP} corresponds to True Positive and FN corresponds to False Negatives

\begin{equation}\label{balance_accuracy}
Balanced Accuracy = \frac{Sensitivity+Specificity}{2}
\end{equation}

\begin{figure*}
    \centering
    \includegraphics[width=0.9\textwidth]{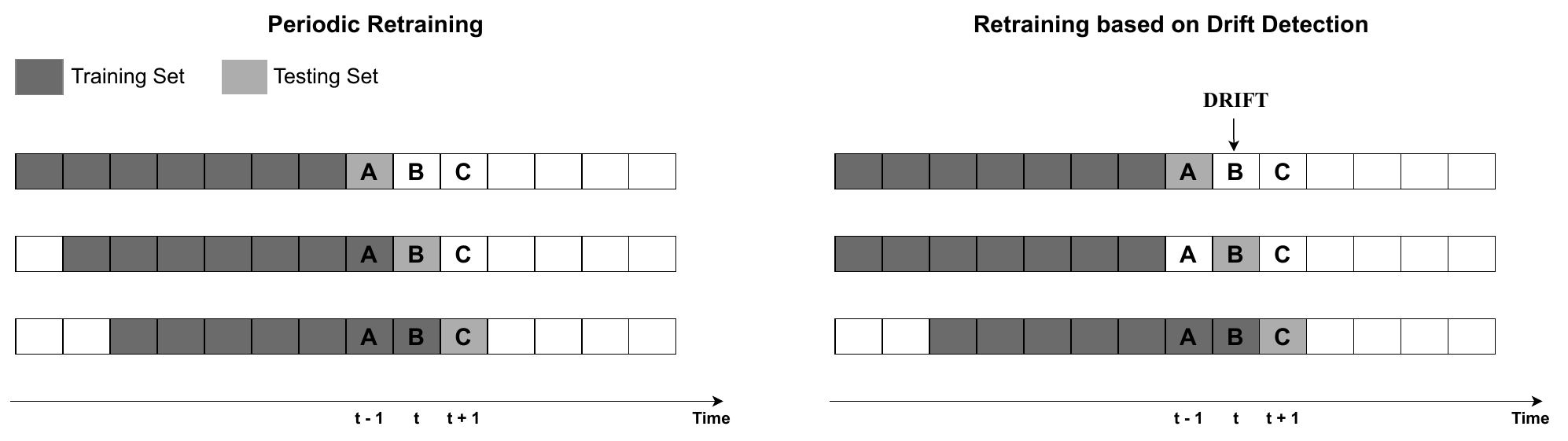}
    \caption{Difference between retraining periodically and retraining based on drift detection.}
    \label{figure:periodic_vs_drift_detection}
    \vspace{-1em}
\end{figure*}

\subsubsection{Evaluation based on Model Performance Preservation}
In this drift detection evaluation technique, we study the effects on the model performance of not updating the model at all (static), updating the model periodically (the most commonly used technique to mitigate the effects of concept drift on AIOps data), and updating the model based on a drift detection technique. By comparing updating a static model with updating a model based on a drift detection technique, we aim to understand whether including a drift detection technique as a retraining indicator significantly increases the performance of a model compared to never updating it. By comparing a periodically updated model with a model updated based on a drift detector we aim to investigate whether the drift detection techniques manage to capture enough drifts to achieve comparable model performance with periodic updates. 

In Figure~\ref{figure:periodic_vs_drift_detection} we depict the difference between periodic retraining and retraining based on drift detection. The former updates a model after each period, while the latter only updates a model when a drift is detected. In our approach, we employ the same sliding-window-based model retraining technique as previous work~\cite{towardsaconsistentinterpretation},~\cite{datasplittingdecisions} where the most recent data is used when retraining.

\subsubsection{Evaluation Label Cost}
\begin{figure}
    \centering
    \includegraphics[width=0.4\textwidth]{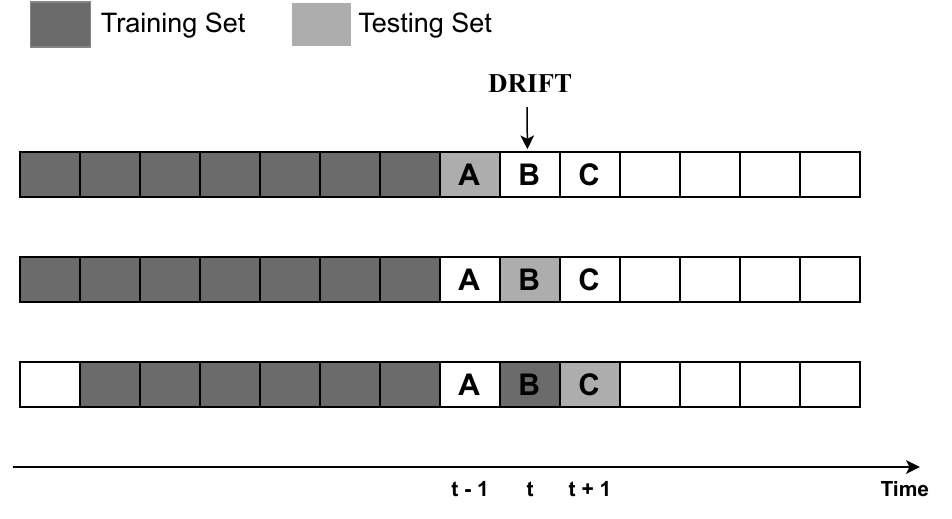}
    \caption{Label cost-efficient maintenance pipeline including McUDI for failure prediction models.}
    \label{figure:cost_label_evaluation}
    \vspace{-1em}
\end{figure}

To understand the benefits of a drift detector in reducing the number of times a model is retrained and redeployed and the cost of obtaining good-quality labels, we propose a new evaluation strategy depicted in Figure~\ref{figure:cost_label_evaluation}. In this evaluation strategy, the retraining is performed solely when drift is indicated by McUDI and only batch B, where the drift was identified, is included in the training set instead of including the most recent data (both batch A and B) as previous work~\cite{datasplittingdecisions},~\cite{towardsaconsistentinterpretation}. If no drift is detected in batch A, the data distribution of batch A is similar to the data distribution in the training data. Therefore, retraining on batch A comes with the cost of obtaining labels for this data without including new information in the training set.

\section{Experiments}
We begin the experimental section by presenting how we extracted the testing batches when concept drift impacts the model performance. We further show our results on the performance of different concept drift indicators in identifying these testing batches. 

\subsection{Extracting the Ground Truth}
To avoid bias in our experiments, we extract the error rate of the employed classifier for 10 random seeds and share them in our open-source package to facilitate reproducibility.

In our experiments, we initially employ the same intervals as previous work~\cite{datasplittingdecisions}, namely one-day periods for the Google dataset and one-month period for the Backblaze dataset. However, while performing the experiments, in most situations drift occurs in all batches of the latter dataset. Since we aim to understand the ability of a drift detector to both identify drift and not raise false alarms, we decreased the size of the period from one month to one week. The reason for this is that from our experiments, the data is significantly different from one month to another and a model trained on the previous month cannot generalize well on the data corresponding to the following month. The same results were not reported in the original work~\cite{datasplittingdecisions}, but this can be a consequence of the difference in data scaling methods.

%\begin{figure}
%    \centering
%    \includegraphics[width=0.5\textwidth]{ground_truth_extraction_results.pdf}
%    \caption{Ground Truth. Batches that contain drift and non-drift for disk and job datasets.}
%    \label{figure:ground_truth_extraction_results}
%    \vspace{-1em}
%\end{figure}

\begin{figure}
     \centering
     \begin{subfigure}[b]{\linewidth}
         \centering
         \includegraphics[width=0.95\textwidth]{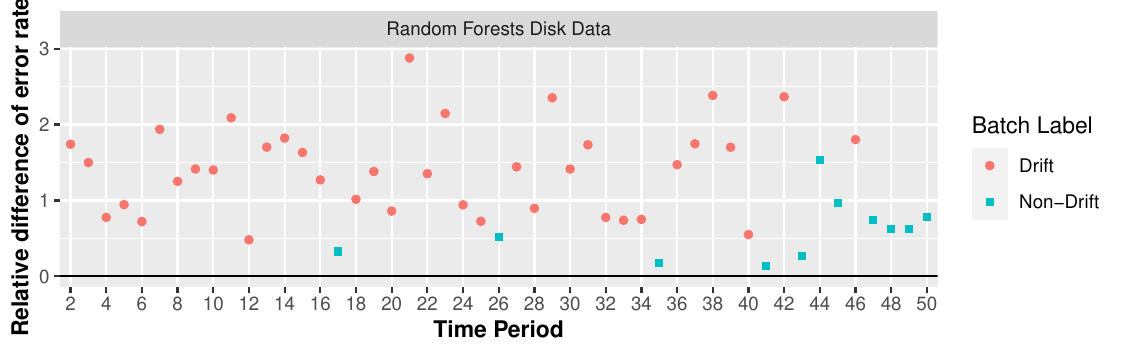}
         %\caption{$y=x$}
         %\label{fig:y equals x}
     \end{subfigure}
     \hfill
     \begin{subfigure}[b]{0.95\linewidth}
         \centering
         \includegraphics[width=\textwidth]{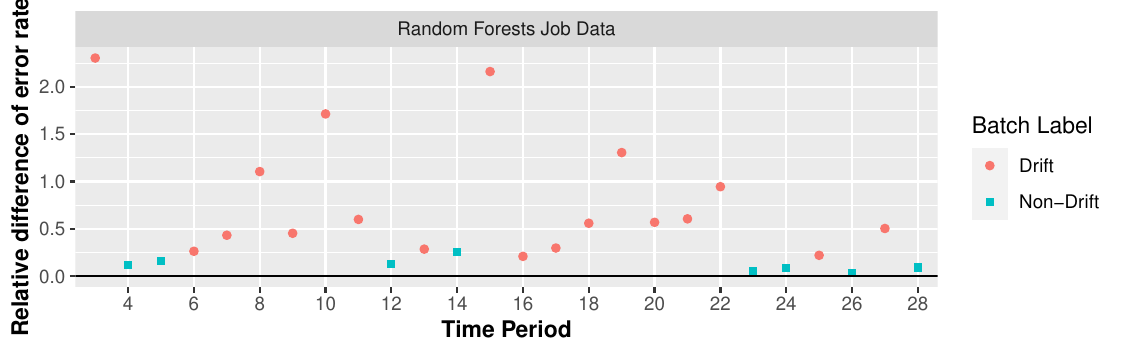}
         %\caption{$y=3\sin x$}
         %\label{fig:three sin x}
     \end{subfigure}
     
        \caption{Ground Truth. Batches that contain drift and non-drift for disk and job datasets.}
        \label{figure:ground_truth_extraction_results}
\end{figure}

Figure \ref{figure:ground_truth_extraction_results} depicts the identified drifts based on the error rate for both datasets and their severity reported as the relative difference of error rate. We consider a data batch status of either drift or non-drift. A drift label is given to batches that were labeled as drift for more than half of the number of random seeds, while the others are labeled as non-drift.

When it comes to the Google dataset, due to high-class imbalance, the data batch corresponding to the first period (first day out of 29 days) contains only samples from the majority class, which makes it impossible for the classifier to learn to distinguish between failure and non-failure. Therefore, we cannot analyze whether there is a concept drift between the first and second periods and we exclude it from the plot. 

\subsection{SoTA Degradation Indicators}
This set of experiments aims to answer the first research question and understand how well state-of-the-art model degradation indicators (number of features that change and changes in data distribution) manage to capture the drifts in the two AIOps datasets. 
\subsubsection{Monitoring Features Individually}

\begin{figure}
     \centering
     \begin{subfigure}[b]{\linewidth}
         \centering
         \includegraphics[width=0.95\textwidth]{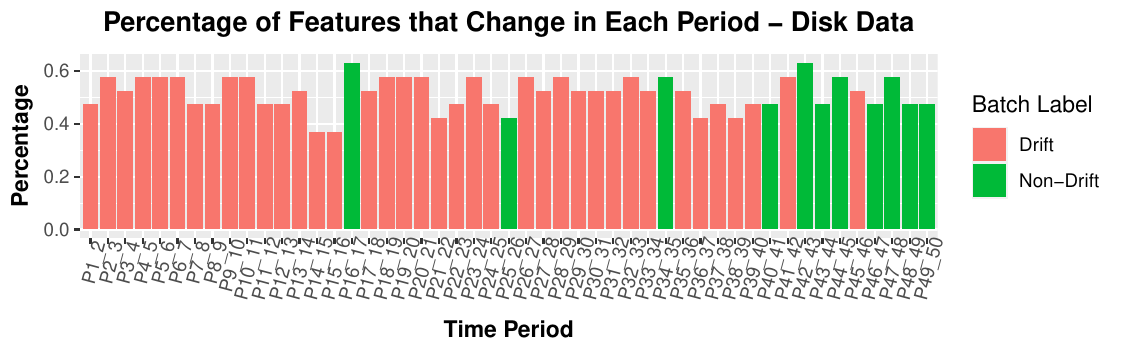}
         %\caption{$y=x$}
         %\label{fig:y equals x}
     \end{subfigure}
     \hfill
     \begin{subfigure}[b]{\linewidth}
         \centering
         \includegraphics[width=0.95\textwidth]{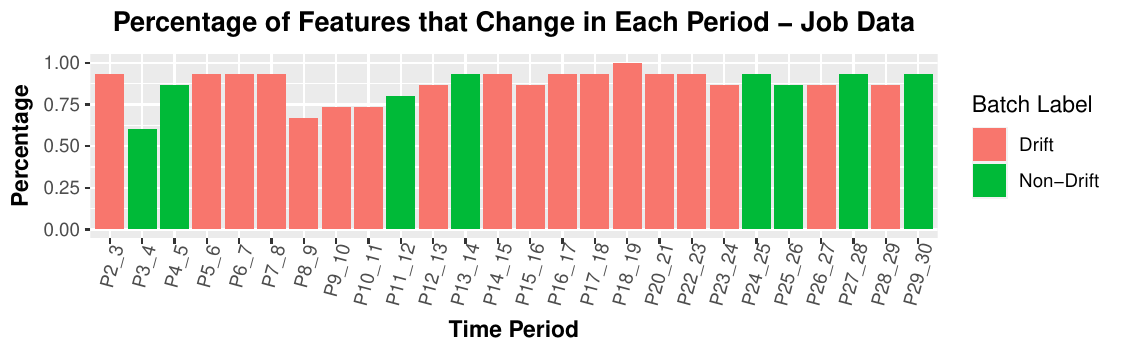}
         %\caption{$y=3\sin x$}
         %\label{fig:three sin x}
     \end{subfigure}
     
        \caption{Percentage of features that change in each period and their corresponding batch label (drift/non-drift).}
        \label{figure:monitoring_features_individually}
\end{figure}

We aim to understand whether the number of features that change is a good indicator of drift. In Figure~\ref{figure:monitoring_features_individually} we display the number of features that are changing in each consecutive two testing batches. It can be noticed that the batches that are not labeled as drift do not necessarily contain fewer features that change, and sometimes fewer features that change are in batches that are labeled as drift. Thus, model degradation is not linked to the number of individual features that change.

%\begin{figure}
%    \centering
%    \includegraphics[width=0.5\textwidth]{monitoring_features_individually.pdf}
%    \caption{Percentage of features that change in each period and their corresponding batch label (drift/non-drift).}
%    \label{figure:monitoring_features_individually}
%    \vspace{-1em}
%\end{figure}

\subsubsection{Monitoring the Distribution of the Data}

%To assess the performance of identifying drifts and non-drifts of a data distribution-based degradation indicator (KS), we compare it with an indicator corresponding to periodic retraining, the SoTA practice to avoid accuracy loss in AIOps solutions over time. We are referring to this degradation indicator as Perio. Since periodic retraining implies retraining after each batch, Perio is a drift indicator that detects drifts in every batch. Thus it should be able to identify all drifts and not identify any non-drifts. 

%In Table~\ref{table:data_distribution_monitoring} we depict the results obtained by a generic model degradation indicator compared to our proposed solution, McUDI. The generic degradation indicator assesses the drift by computing the differences in the distribution of all features between training and testing, while McUDI computes the distribution from solely the most important features.

To understand how accurately KS detects drifts, we compute the three aforementioned accuracy metrics and display the results in Table~\ref{table:data_distribution_monitoring_mcudi}. We can observe that the sensitivity is close to 0.0 for both job and disk datasets, which shows that KS does not manage to capture non-drifts. However, we can see that KS identifies plenty of drifts, especially on the job dataset, where the specificity is 90\%. Thus, KS can accurately capture drift with the cost of raising plenty of false alarms.

%From Table~\ref{table:data_distribution_monitoring} we can notice that while the generic approach captures more drifts, it tends to label every batch as a drift. This can be observed through the poor performance this technique has when detecting non-drift batches. Thus, one strong advantage of McUDI is that it is less sensitive to changes in the data that are not impacting the model's performance.

\subsection{Model-Centric Degradation Indicator}
With this experiment, we answer our second research question and investigate the performance of a model-centric degradation indicator (McUDI) compared to SoTA practices. We evaluate McUDI based on the drift detection accuracy (RQ2a) and the performance preservation over time (RQ2b).

%a degradation indicator corresponding to periodically updating a failure prediction model (Periodic) and to a degradation indicator that is not model-centric (KS). 

\subsubsection{Evaluation based on Drift Detection Accuracy}
We evaluate the drift detection performance of McUDI in comparison with a degradation indicator corresponding to periodically updating a failure prediction model (Periodic) and to the degradation indicator that is not model-centric (KS). 

\begin{table}[ht]
\centering
\caption{Drift detection accuracy metrics for Periodic, KS, and McUDI. With \textbf{bold} we depict the highest value for each metric for each dataset.}
\begin{tabular}{| c | c | r | r | r | }

 \cline{2-5}
 \multicolumn{1}{c}{\textbf{}} &
  \multicolumn{1}{|c}{\textbf{Metric}} &

 \multicolumn{1}{|c|}{\textbf{Periodic}} &
 \multicolumn{1}{|c}{\textbf{KS}} &
 \multicolumn{1}{|c|}{\textbf{McUDI}} \\
 \cline{1-5}

\multirow{3}{0.1em}{\rotatebox[origin=c]{90}{\textbf{Disk}}}

 & Balanced Accuracy & 0.5 & 0.41 & \textbf{0.57} \\
 & Specificity & \textbf{1.0} & 0.72 &  0.69 \\
 & Sensitivity & 0.0 & 0.09 & \textbf{0.44} \\
 
  \cline{1-5}

\multirow{3}{0.1em}{\rotatebox[origin=c]{90}{\textbf{Job}}} 
& Balanced Accuracy & 0.5 & \textbf{0.45} & 0.42  \\
& Specificity & \textbf{1.0} & 0.90 & 0.82 \\
& Sensitivity & 0.0 & 0.00 & \textbf{0.01}\\
 
 \cline{1-5}

\end{tabular}
\label{table:data_distribution_monitoring_mcudi}
\end{table}

We show our results in Table~\ref{table:data_distribution_monitoring_mcudi}. The Periodic degradation indicator identifies all drifts (specificity of 1.0) and does not identify any non-drifts (sensitivity of 0.0). It can be noticed from Table~\ref{table:data_distribution_monitoring_mcudi} that the sensitivity of KS is similar to the one of Periodic for disk data (0\%), suggesting that the KS degradation indicator closely resembles periodic model retraining. 

On the other side, McUDI achieves the highest sensitivity for both datasets, showing its ability to accurately identify non-drifts and raise fewer false alarms compared to Periodic and KS. The highest sensitivity obtained by McUDI is 0.44 for disk data. In a more rigorous analysis, we noticed that for the disk dataset, the Random Forest classifier considers around 7 out of 19 features as important based on the mean importance threshold that is used within McUDI. This may indicate that the high number of non-important features that change severely impact the data distribution and can justify the high number of false alarms raised by KS and why McUDI is beneficial for this dataset. When it comes to the job dataset, around 9 out of 15 features are considered important, which can indicate that changes in non-important features do not considerably impact the data distribution. Therefore, we demonstrated that McUDI achieves good results in properly detecting non-drifts when the classifier considers less than half of the employed features as important. Furthermore, McUDI identifies more than 65\% of the occurring drifts for both datasets, obtaining a specificity of around 69\% and 82\% for disk and job data, respectively.

\subsubsection{Evaluation based on Performance Preservation}
With this experiment, we aim to understand the effects of having a drift detector as an indicator of when a model should be retrained. Thus, we compare the effect of employing McUDI and KS in the maintenance pipeline of AIOps failure prediction models. We also report the performance of never retraining the model (Static) and the performance of retraining the model periodically (Periodic) as lower and upper baselines, respectively. We further report the number of times retraining is necessary with respect to the total number of possible retraining times averaged over the 10 random seeds.

\begin{table}[ht]
\centering
\caption{Performance of each retraining strategy with respect to the number of required retraining times. \textit{ROCAUC} is the metric we use to assess the performance of the failure prediction model. \textit{Drifts} refers to the number of identified drifts/number of total testing batches.}
\begin{tabular}{| c | r | r | r | r | }

 \cline{2-5}
  \multicolumn{1}{c}{\textbf{}} &

 \multicolumn{2}{|c|}{\textbf{Disk Data}} &
 \multicolumn{2}{|c|}{\textbf{Job Data}}\\

 \cline{1-5}
  \multicolumn{1}{|c}{\textbf{Model}} &

 \multicolumn{1}{|c|}{\textbf{ROCAUC}} &
 \multicolumn{1}{|c|}{\textbf{Drifts}} &
 \multicolumn{1}{|c|}{\textbf{ROCAUC}}&
 \multicolumn{1}{|c|}{\textbf{Drifts}}\\
 \cline{1-5}

  Static & 0.90 & 0/25 & 0.83 & 0/14 \\

  \cmidrule{1-5}
  KS & 0.91 & 25/25 & 0.86 & 11.9/14\\
  McUDI & 0.91 & 23.4/25 & 0.86 & 12.5/14\\
  \cmidrule{1-5}
  Periodic & 0.91 & 25/25 & 0.87 & 14/14\\

 \cline{1-5}

 \cline{1-5}
\end{tabular}
\label{table:performance_preservation_ks_mcudi}
\end{table}

We depict our results in Table~\ref{table:performance_preservation_ks_mcudi}. The lower baseline, Static achieves the lowest performance in terms of ROC AUC and requires 0/25 retraining times. The upper baseline, Periodic, achieves the highest ROC AUC and requires constant retraining (25/25). From Table~\ref{table:performance_preservation_ks_mcudi} we can notice that, in case of disk data, Periodic, KS, and McUDI achieve similar results in terms of ROC AUC (around 91\%), while McUDI requires the lowest number of retraining times (23.5). We can again observe that the performance of KS and Periodic is identical. When it comes to the job dataset, both McUDI and KS achieve similar performance in terms of ROC AUC to Periodic with a lower number of required retraining times.

\subsection{Label Costs Evaluation}

In this experiment, we answer our third research question and show the effect of including a McUDI in the maintenance pipeline of AIOps solutions in terms of label costs. In other words, we compare the label costs when performing periodic retraining with the label costs when employing McUDI, as well as the performance in terms of average ROC AUC when employing each of the two types of retraining. We quantify label costs as the number of samples that require annotation to retrain the AIOps model.

\begin{table}[ht]
\centering
\caption{Results of label costs assessment. \textit{ROCAUC} is the metric we use the assess the performance of the failure prediction model. \textit{L Cost} refers to the number of necessary samples that require expert annotation to perform retraining.}
\begin{tabular}{| c | r | r | r | r | } 

 \cline{2-5}
  \multicolumn{1}{c}{\textbf{}} &

 \multicolumn{2}{|c|}{\textbf{Disk Data}} &
 \multicolumn{2}{|c|}{\textbf{Job Data}}\\
 
 \cline{2-5}
 \multicolumn{1}{c}{\textbf{}} &

 \multicolumn{1}{|c|}{\textbf{ROCAUC}} &
 \multicolumn{1}{|c|}{\textbf{L Cost}}&
 \multicolumn{1}{|c|}{\textbf{ROCAUC}}&
 \multicolumn{1}{|c|}{\textbf{L Cost}}\\

 \cline{1-5}

 McUDI & 0.91 & 3,745,919 & 0.86 &  304,326\\
 Periodic & 0.91 & 4,005,978 & 0.87 & 333,982\\

 \cline{1-5}

\end{tabular}
\label{table:label_costs}
\end{table}

From Table~\ref{table:label_costs} we  notice that the performance of McUDI is almost similar to the one obtained by periodically retraining models for both disk and job data. The advantage of employing McUDI is that it reduces the number of necessary retraining times as well as the required labels to perform the retraining. Thus, by employing our proposed maintenance pipeline, the number of samples that need annotation is reduced by approximately 260k and 30k samples for disk and job data respectively.

%From the previous experiments, we noticed that monitoring techniques that assess the differences in distribution over time are capable of properly capturing the drifts that lead to performance degradation. Thus, in the following experiments, we aim to answer our second research question and understand whether computing the data distribution from only the most important features according to the model improves drift identification. Thus, in Table~\ref{table:data_distribution_monitoring_mcudi} we show the performance of McUDI against periodic retraining

\section{Discussions}

%\textbf{TODO}
Throughout our experiments, we showcase that one commonly used unsupervised technique to monitor concept drift, namely the number of features that change, is not a good performance degradation indicator for failure prediction AIOps models. However, monitoring the distribution manages to accurately capture the moments when concept drift affects the accuracy of the AIOps solution. Thus, we encourage the AIOps community to constantly monitor changes in the data distribution over time as a model quality indicator.

While experimenting with both McUDI and KS, we noticed the benefits of our solution on the disk dataset. McUDI raises significantly less false alarms when identifying drifts when predicting disk failure compared to KS. This can be explained by the fact that the number of features that the classifier considers important is almost half of the total features employed. Thus, when computing the data distribution using all features, the KS statistical test might detect drift when non-important features are changing, but they don't impact the performance of the model. This makes a model centric approach such as McUDI an accurate indicator of model degradation with strong evidence on disk failure AIOps solutions.

%While monitoring changes in the data distribution over time, we employed both KS a commonly use unsupervised drift detection technique, and our proposed solution, McUDI, a model-centric unsupervised drift detection technique. We noticed that KS raises plenty of false alarms compared to McUDI while obtaining similar performance in terms of accuracy. This shows that the same model accuracy can be obtained if the model is retrained less, making McUDI a better indicator of when model retraining is necessary. Therefore, we urge the drift detection community to focus on developing more model-centric unsupervised drift detection methods that can be used to monitor the quality of failure prediction AIOps solutions. 

Besides the constant retraining and model redeployment associated with current failure prediction AIOps model retraining practices (periodic retraining), there is also a high number of high-quality labels that need to be provided by domain experts periodically. In our work, we provide a model maintenance pipeline, which employs McUDI as a technique to verify when the model requires retraining. By employing our maintenance pipeline instead of periodical retraining, AIOps researchers and practitioners can lower both the times the model needs to be redeployed and the number of high-quality labels provided by domain experts, which saves annotation costs. Therefore, we recommend employing this model maintenance pipeline for AIOps applications integrated into larger systems for which redeployment is expensive or for AIOps applications for which obtaining the labels comes with a substantial demand from experts for manual data annotation

\section{Conclusions}

In this paper, we propose McUDI a novel model-centric unsupervised performance degradation indicator for classification problems that can identify when failure prediction AIOps models require retraining. McUDI is unsupervised since it does not require labels to identify changes in data that lead to model performance degradation. Furthermore, McUDI is model-centric since it checks for data changes solely among the most important features indicated by the trained model.

%does not take all the available features in the dataset into account, but it only considers the features that are important to the trained model. 

We evaluate the performance of McUDI on two open-source operational datasets used for building failure prediction AIOps models. We empirically prove that McUDI can be beneficial when included in the maintenance pipeline of AIOps solutions since it lowers the number of times the models need to be retrained and redeployed while preserving their performance compared to the current state-of-the-art practice of periodically retraining the model. We demonstrate that McUDI surpasses KS, a commonly used unsupervised technique to detect data changes, in terms of raising false alarms. Moreover, we propose a maintenance pipeline for failure prediction AIOps solutions that lowers both the number of retraining times and the cost of getting high-quality labels.

In the future, we aim to evaluate McUDI on new operational private datasets and estimate the saved costs of including McUDI in the maintenance pipeline of existing AIOps solutions. Furthermore, McUDI will be assessed with other tree-based classifiers for disk and job failure prediction. Moreover, although solely assessed on operational data in this work, McUDI should be applicable for other types of datasets and, thus, future work can consider evaluating McUDI on datasets belonging to other domains than AIOps.

\appendix

%\section*{Ethical Statement}

%There are no ethical issues.

\bibliographystyle{named}
\bibliography{ijcai23}

\end{document}